\documentclass{PoS}
\usepackage[T1]{fontenc}
\usepackage[utf8]{inputenc}
\usepackage{amsmath,amssymb,amsfonts}
\usepackage{multicol}
\usepackage{subfigure}
\usepackage{float}

\title{Numerical Stochastic Perturbation Theory and  Gradient Flow in $\varphi^4$ Theory}

\ShortTitle{NSPT in $\varphi^4$}

\author{Mattia Dalla Brida \\
        NIC, DESY, Platanenallee 6, 15738 Zeuthen, Germany\\
        E-mail: \email{mattia.dalla.brida@desy.de}}
        
\author{\speaker{Marco Garofalo}\\
	 Higgs Centre for Theoretical Physics, School of Physics and Astronomy, 
	 \\ The University of Edinburgh, Edinburgh EH9 3JZ, Scotland, UK\\
        E-mail: \email{M.Garofalo@sms.ed.ac.uk}}
     
\author{Anthony D. Kennedy\\
	 Higgs Centre for Theoretical Physics, School of Physics and Astronomy, 
	 \\ The University of Edinburgh, Edinburgh EH9 3JZ, Scotland, UK\\
        E-mail: \email{adk@ph.ed.ac.uk}}

\abstract{ In this contribution we present an exploratory study of several novel methods
	   for numerical stochastic perturbation theory. For the investigation we consider
	   observables defined through the gradient flow in the simple $\varphi^4$ theory.}

\FullConference{The 33rd International Symposium on Lattice Field Theory\\
		14 -18 July 2015\\
		Kobe International Conference Center, Kobe, Japan}

\begin{document}

\section{Introduction}
\begin{samepage}
The information from lattice perturbation theory (LPT) can be valuable for 
non-perturbative investigations of lattice field theories. For instance LPT allows 
the matching of renormalization schemes at high-energy to be determined,
and it can provide useful insights on the lattice artifacts of the observables of 
interest. As  well known, LPT is made more difficult than its continuum
counterpart by the complicated expressions for vertices and propagators that normally
force numerical evaluation even for simple diagrams. 
Additionally, in the case of gauge theories the appearance of new vertices at all orders
of perturbation theory makes the number of diagrams grow rapidly with the perturbative
order, thus leaving only low-order results accessible to standard techniques.
Numerical stochastic perturbation theory (NSPT) was introduced some time 
ago~\cite{DiRenzo:1994sy} (see~\cite{Hesse:2013lat,Brida:2013mva} for
recent developments) in order to bypass these difficulties, and thus allow  
estimates of high-order perturbative coefficients to be obtained. 
The basic idea of NSPT is to integrate numerically a discrete
version of the equations of stochastic perturbation theory. 
The method comes with two main limitations: first of all it is
not exact, a sequence of simulations with a finer and finer discretization of the relevant 
equations have to be performed in order to extrapolate away the systematic errors in the
results, secondly, the numerical simulations suffer from critical slowing down as the 
continuum limit of the theory is approached.

Martin L\"uscher has recently proposed a new form of stochastic perturbation
theory, namely Instantaneous Stochastic Perturbation Theory (ISPT)~\cite{Luscher:2014mka}, 
which completely eliminates these limitations; this contribution 
presents an exploratory study of this technique. Moreover, we propose new NSPT 
methods using stochastic equations other than the Langevin equation. As we shall 
see this leads to more efficient numerical algorithms that can 
significantly alleviate the limitations of the standard set-up. For simplicity we 
test these methods in a simple scalar $\varphi^4$ theory defined on a finite Euclidean
4-dimensional periodic lattice of size $L$ with action given by,
  \begin{equation}
  \label{eq:S}
   S(\varphi)=
  \sum_x\,\left\{\frac{1}{2}\partial_{\mu}\varphi(x)\partial_{\mu}\varphi(x) + 
  \frac{1}{2} m_0^2\varphi(x)^2+{\frac{g_0}{4!}}\varphi(x)^4\right\}.
\end{equation}
Here $\partial_\mu \varphi(x)=[\varphi(x+a\hat\mu)-\varphi(x)]/a$, with $\mu=0,\ldots,3$ 
defines the usual forward lattice derivative where $\hat\mu$ is the unit vector in the 
direction $\mu$ and $a$ is the lattice spacing. The parameters $m_0$ and $g_0$ 
are the bare mass and coupling, respectively.

\section{ISPT}

ISPT is based on the concept of trivializing maps~\cite{Luscher:2014mka}. These transform 
Gaussian distributed fields $\eta_i(x)$ into a stochastic field $\phi(x)$
such that,
\begin{equation}
  \label{eq:Trivial}
  \langle \phi(x_1)...\phi(x_n) \rangle_\eta=\langle \varphi(x_1)...\varphi(x_n) \rangle,
\end{equation}
where $\varphi$ is the field with action (\ref{eq:S}), $\langle\cdots\rangle$
is the expectation value in the $\varphi^4$ theory, and $\langle\cdots\rangle_\eta$
denotes the average over the fields $\eta_i(x)$ which satisfy
\begin{equation}
 \langle \eta_i(x)\eta_j(y)\rangle_\eta= \delta_{ij}\,\delta_{xy},\qquad
 \langle\eta_i(x)\rangle_\eta=0.
\end{equation}
Without entering into the details (for which we refer to~\cite{Luscher:2014mka}), 
the stochastic field $\phi(x)$ can be written as a power series in the couplings of 
the theory, whose coefficients are calculable functions of the noise fields $\eta_i(x)$.                                                           
For $\varphi^4$ theory\end{samepage}  
\begin{equation}
  \label{eq:PowerSeries}
  \phi(x)=\sum_{j,k=0}^{\infty}\phi_{(j,k)}(x)(\delta m^2)^{j}  g_0^k,
\end{equation}
where the coefficients $\phi_{(j,k)}(x)$ depend on the values of the noise fields
$\eta_i(x)$. Note that the expansion is in terms of the bare coupling $g_0$,
and the mass counterterm $\delta m^2=m_0^2-m^2$, where $m$ is the renormalized mass. 
This method generates uncorrelated field configurations, unlike methods based on Markov 
processes (see below), and it is exact in the sense that the only source of uncertainly 
in the results comes from the stochastic evaluation of the Gaussian integrals in 
(\ref{eq:Trivial}). Being a diagrammatic technique, however, the number of
contributions and thus the cost of evaluating the coefficients $\phi_{(j,k)}$ grows 
rapidly with the perturbative order~\cite{Luscher:2014mka}. 

For this work we wrote a code for the automated computation of the trivializing 
map $\phi$ in the $\varphi^4$ theory to arbitrary order in the couplings.\footnote{For 
the construction of the trivializing map we used the publicly available code 
provided by Martin L\"uscher at \texttt{luscher.web.cern.ch/luscher/ISPT}.} As a 
test of our implementation we computed the finite-volume coupling defined 
in~\cite{Weisz:2010xx} for several lattice volumes and mass values, and compared the 
results with their analytic perturbative expansion up to three-loop order.
We found excellent agreement within the per-mill precision we reached.

\section{LSPT}

The original way to implement numerical stochastic perturbation theory~\cite{DiRenzo:1994sy}
is to generate the stochastic field $\phi(x)$ via a Markov process based on the Langevin 
equation (LSPT),
\begin{equation}
  \label{eq:Langevin}
  \partial_{t_s} \phi(x,t_s)=\partial^2\phi(x,t_s)
  -(m^2+\delta m^2) \phi(x,t_s) - \frac{g_0}{3!} \phi(x,t_s)^3+\eta(x,t_s),
\end{equation}
where $t_s$ is the simulation time, $\partial^2$ the lattice Laplacian, and $\eta(x,t_s)$ 
is a Gaussian distributed noise field satisfying $\langle \eta(x,t_s)\rangle_\eta=0$ and
$\langle \eta(x,t_s)\eta(y,t'_s)\rangle_\eta=2 \delta_{xy}\delta(t_s-t'_s)$.
As in ISPT, the field $\phi(x,t_s)$ is assumed to have an expansion of the form 
(\ref{eq:PowerSeries}), while the noise $\eta(x,t_s)$ only has a lowest order component.
The expectation values of the target theory are then obtained as,
\begin{equation}
 \label{eq:Equilibrium}
 \langle\phi(x_1,t_s)...\phi(x_n,t_s) \rangle_\eta \overset{t_s\to\infty}{=}
 \langle \varphi(x_1)...\varphi(x_n) \rangle.
\end{equation}
The way one proceeds is to discretize the stochastic time as $t_s=n\epsilon$, with 
$\epsilon$ the step-size, and integrate the Langevin equations numerically order-by-order
in couplings $g_0$ and $\delta m^2$ according to a given integration scheme.
Consequently, one expects step-size errors to effect the results, which thus need to be 
extrapolated away. Moreover, the fields $\phi(x,t_s)$ generated in this way are correlated
to each other. On the other hand, the cost of the method scales only quadratically with the 
perturbative orders in the couplings.

\section{HSPT}

Taking inspiration from the Langevin approach, we also considered whether using different 
stochastic differential equations might improve LSPT. 

One possibility is to use the Hybrid Molecular Dynamics equations (HSPT),
\begin{gather}
  \label{eq:HMD}
  \partial_{t_s}\phi(x,t_s)=\pi(x,t_s),\quad\quad
  \partial_{t_s}\pi(x,t_s)=\partial^2\phi(x,t_s)
  -(m^2+\delta m^2) \phi(x,t_s) - \frac{g_0}{3!} \phi(x,t_s)^3.
 \end{gather}
Here the momentum field $\pi(x,t_s)$ is also considered as a series 
(\ref{eq:PowerSeries}) in the couplings, and it is sampled from a Gaussian distribution
$P(\pi)\propto e^{-\frac{\pi^2}{2}}$ at the beginning of each trajectory. In particular,
at the start of each trajectory the momenta only have a non-zero lowest order component, 
while they acquire higher-order components through the MD evolution. As in the Langevin
case, the simulation time needs to be discretized, $t_s=n\delta t$, and the MD
equations integrated numerically order-by-order in $g_0$ and $\delta m^2$ up to a time 
$t_s=\tau$ according to some integrator. Expectation values in the target
theory are then similarly obtained as in (\ref{eq:Equilibrium}) by averaging over the 
trajectories~\cite{Luscher:2011qa}. In order to have an ergodic algorithm, one
needs to randomize the trajectory lengths $\tau$ so to update all frequency components of the 
field $\phi$, including the ones of the lowest order field $\phi_{(0,0)}$~\cite{Mackenzie:1989us}.
 
\section{KSPT}

Another possibility that is worth exploring is to generate the field $\phi(x)$ through 
a stochastic evolution in phase-space according to Kramers equations~\cite{Horowitz:1986dt}
(see also~\cite{Luscher:2011qa}) (KSPT),
\begin{equation}
 \label{eq:Kramers}
 \begin{split}
  \partial_{t_s}\phi(x,t_s)&=\pi(x,t_s),\\
  \partial_{t_s}\pi(x,t_s)&=-\gamma\pi(x,t_s) + \partial^2\phi(x,t_s)-(m^2+\delta m^2) 
  \phi(x,t_s) - \frac{g_0}{3!} \phi(x,t_s)^3 + \eta(x,t_s).  
 \end{split}
\end{equation}
The corresponding algorithm is obtained by alternating a partial refreshment of the 
momentum field,
$
 \pi'=e^{-\gamma\delta t}\pi+\sqrt{1-e^{-2\gamma\delta t}}\eta,
$
with the numerical integration of the MD equations for a time-step,
using a suitable integrator.
Here $\eta(x,t_s)$ is a Gaussian noise field with zero mean and variance
$\langle\eta(x,t_s)\eta(y,t'_s) \rangle=\delta_{xy}\delta(t_s-t'_s)$,
while $\gamma$ is a free parameter that may be tuned to minimize autocorrelations.
As in the Langevin case, the noise field only has a lowest order component, while 
the fields $\phi$ and $\pi$ have an expansion of the form (\ref{eq:PowerSeries}). 
For $\gamma=0$ the algorithm reduces to a single-step HSPT with 
trajectory length $\tau=\delta t$. On the contrary, if the continuum limit of the 
theory is taken keeping the parameter $\gamma$ fixed in lattice units, the equations
(\ref{eq:Kramers}) can be shown to reduce to the Langevin equation 
(\ref{eq:Langevin})~\cite{Luscher:2011qa}.

\section{Observables}

In order to test the different methods we rely on observables defined through the 
gradient flow~\cite{Luscher:2010iy}. This allows us to obtain simple and precise 
quantities with a well defined continuum limit. In the case of the $\varphi^4$ theory 
the gradient flow equations can be defined as~\cite{Monahan:2015lha}:
$ \partial_t \tilde{\varphi}(x,t)=\partial^2\tilde{\varphi}(x,t)$, 
$\tilde{\varphi}(x,0)=Z^{-1/2}_\varphi\varphi(x)$, where $t\geq 0$ is the flow time, 
and $Z_\varphi$ is the wavefunction renormalization. We then consider the dimensionless
quantity $\mathcal{E}=t^2 \langle E \rangle$, where $E$ is given by the quartic energy 
density of the field at positive flow time, i.e.,
$\langle E \rangle= \langle \tilde{\varphi}(x,t)^4 \rangle= 
E_0+ E_1 g_0+ E_2g_0^2+ E_3g_0^3 +{\rm O}(g_0^4).$
In particular, we are interested in studying the continuum limit of $\mathcal{E}$
keeping the box size $L$, the mass $m$, and the flow time $t$, fixed in physical units. 
To this end, we introduce the dimensionless constants,
$z=mL$ and $ c=\sqrt{8t}/L.$ The continuum limit is then taken by increasing the lattice
size $L/a$ and decreasing the lattice mass $am$, while holding $z$ and $c$ fixed. 

The renormalization of the mass can be fixed by requiring~\cite{Weisz:2010xx},
\begin{equation}
  \label{eq:MassRen}
  \frac{\chi_2}{\chi_2^*}=\left( 1+\frac{\hat p_*^2}{m^2}\right),\qquad p_*=(2\pi/L,0,0,0), 
\end{equation}
where $\hat p^2=\sum_\mu \hat{p}_\mu^2$ with $\hat p_\mu=2\sin(p_\mu/2)$, and 
$p_\mu$ are the lattice momenta in a periodic box. $\chi_2$ and $\chi_2^*$, 
are the connected two-point functions of the field $\varphi(x)$ evaluated at momentum 
$p=0$ and $p_*$, respectively. Once the relation (\ref{eq:MassRen}) is computed as a 
power series in both $\delta m^2$ and $g_0$, this condition allows us to determine 
$\delta m^2$ as a series in $g_0$. Finally, the 
wavefunction renormalization of the bare fields at $t=0$ can be defined as: 
$Z^{-1}_\varphi = (\chi_2^{*\,-1} - \chi_2^{-1} )/\hat{p}^2_*$~\cite{Weisz:2010xx}.

\section{Results}

\begin{figure}[!hpbt]
\includegraphics[width=0.50\textwidth]{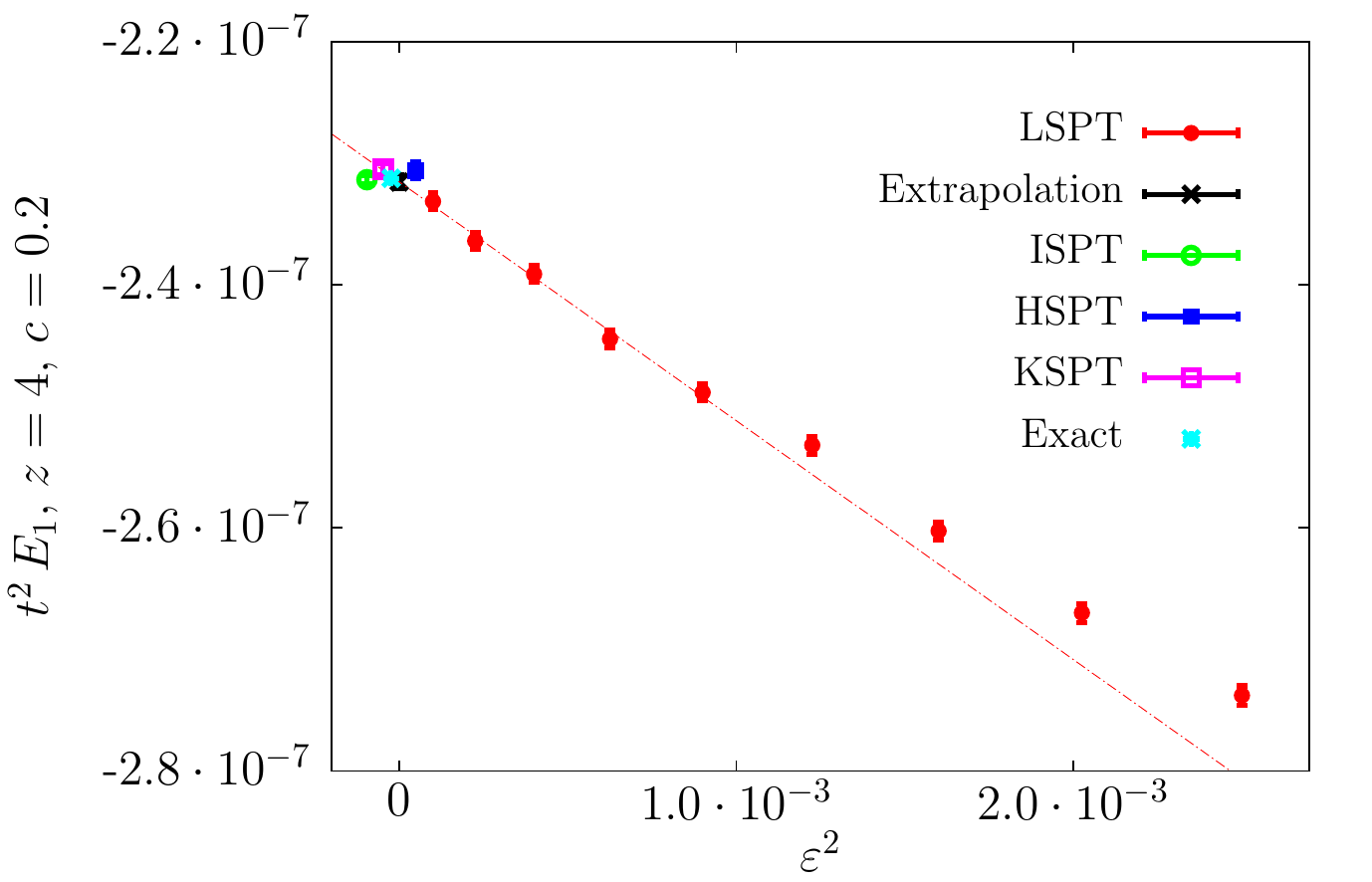}
\includegraphics[width=0.50\textwidth]{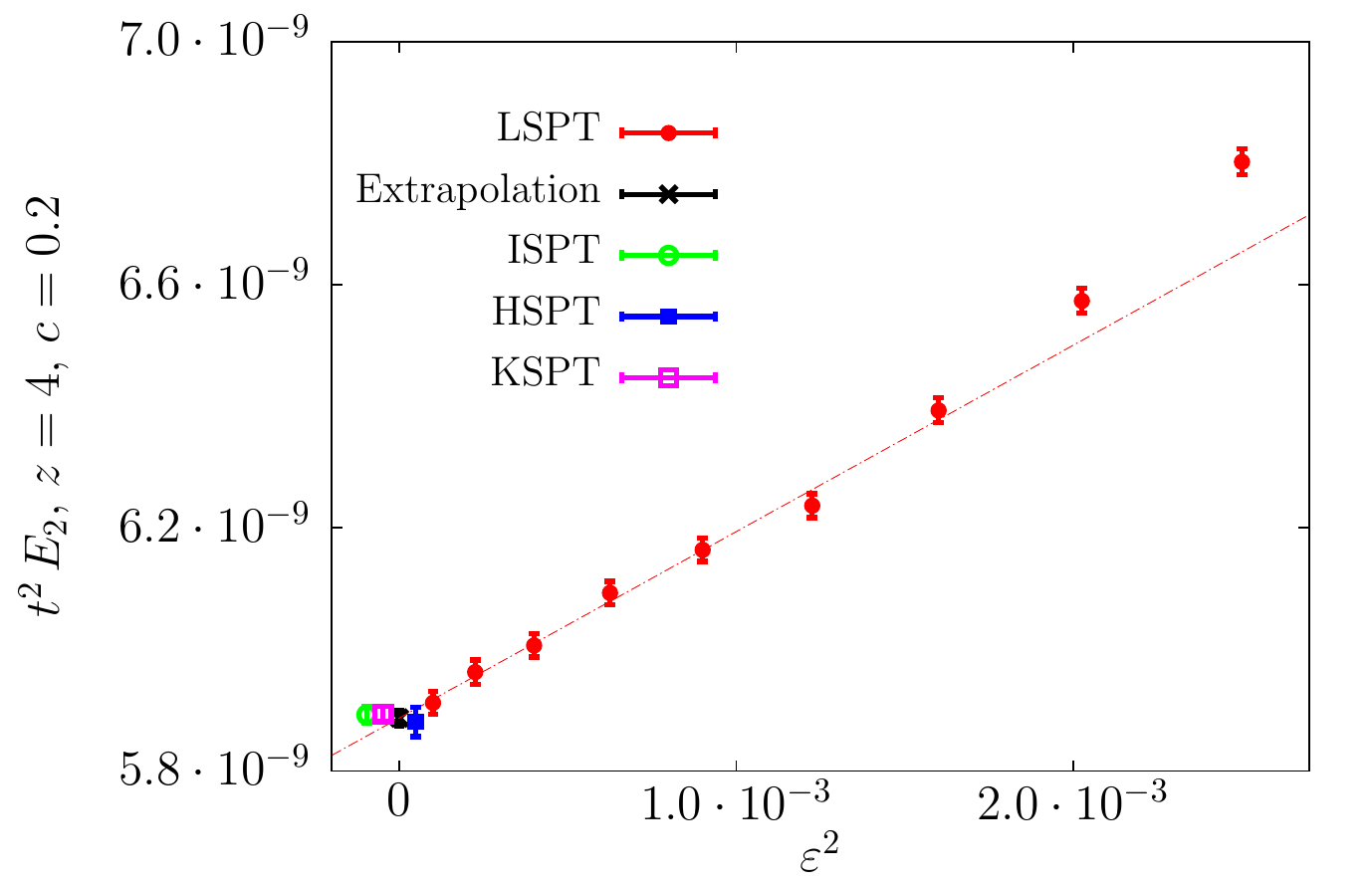}
\caption{Comparison of different methods in the determination of $t^2E_1$ and $t^2E_2$
	 for $z=4$, $c=0.2$, and $L/a=4$. The analytic 
	 result, ``Exact'', the result of the extrapolation for LSPT, ``Extrapolation'', 
	 as well as the ISPT, KSPT and HSPT results are plotted near $\epsilon=0$. }
\label{fig:Comparison}
\end{figure}

In Figure \ref{fig:Comparison} we present a check on the consistency
of the different methods in the perturbative computation of $\mathcal{E}$. The extrapolations
are compared with analytic results where available. 

In the figure we show the results for LSPT for different values of the 
step-size $\epsilon$, and corresponding extrapolations $\epsilon\to0$. 
The integration of the Langevin equations has been performed using a $2^{\rm nd}$ order
Runge-Kutta scheme: we therefore expect O($\epsilon^2$) errors in the observables. The results
for KSPT and HSPT, have been obtained using a $4^{\rm th}$ order symplectic 
integrator for the MD equations with step-size $\delta t=0.5$. This is expected to introduce 
O($\delta t^4$) corrections. Given our choices of integrators the results from HSPT, and KSPT 
show no sign of step-size errors within the statistical accuracy if compared with the
analytic results or ISPT. In the case of LSPT, instead, the step-size errors are significant
even though the step-sizes considered all satisfy, $\epsilon^2 < \delta t^4$; nevertheless the
results agree with the other methods after extrapolation to $\epsilon\to0$.

\begin{figure}[!hptb]
\includegraphics[width=0.50\textwidth]{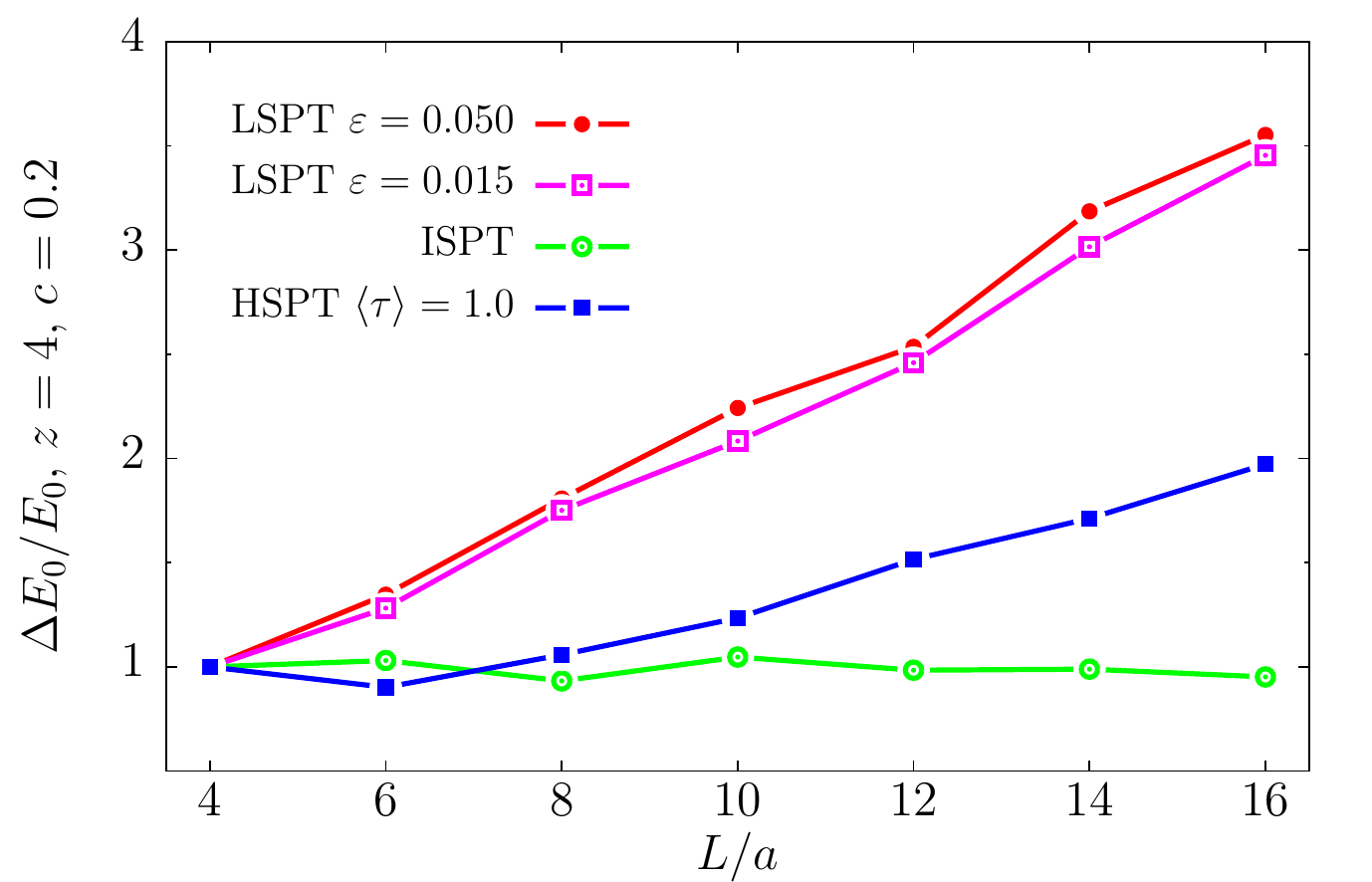}
\includegraphics[width=0.50\textwidth]{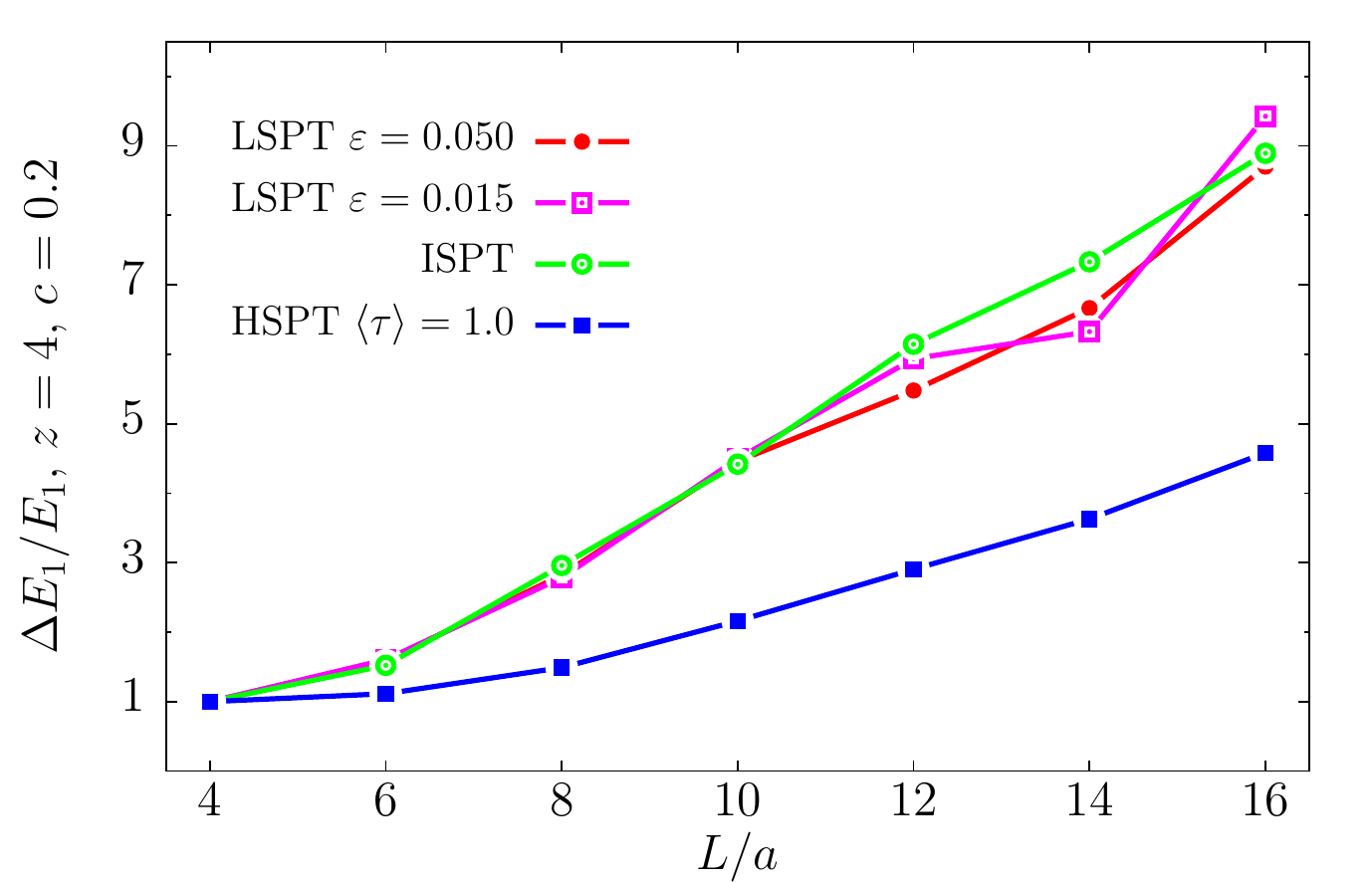}
\includegraphics[width=0.50\textwidth]{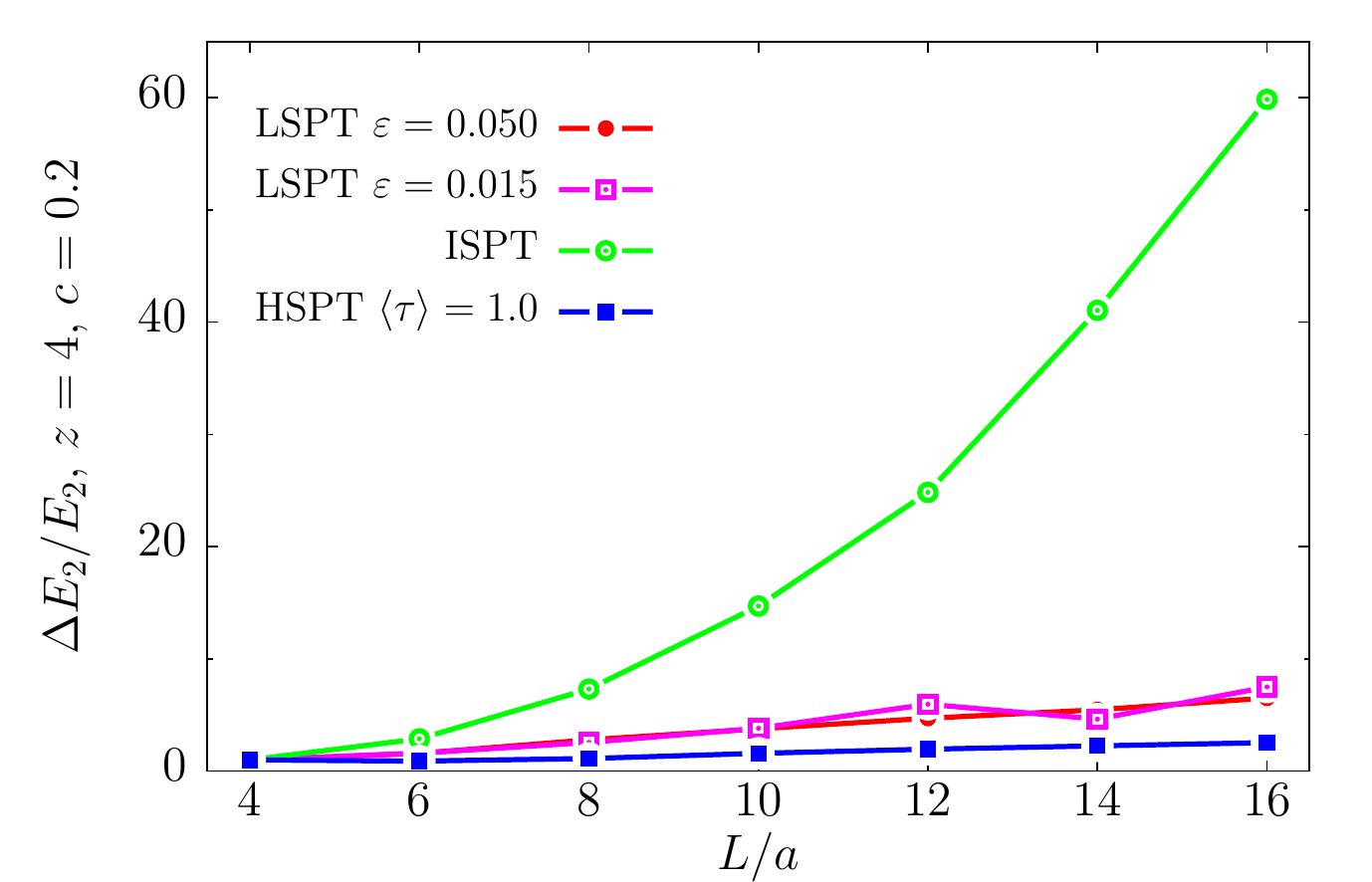}
\includegraphics[width=0.50\textwidth]{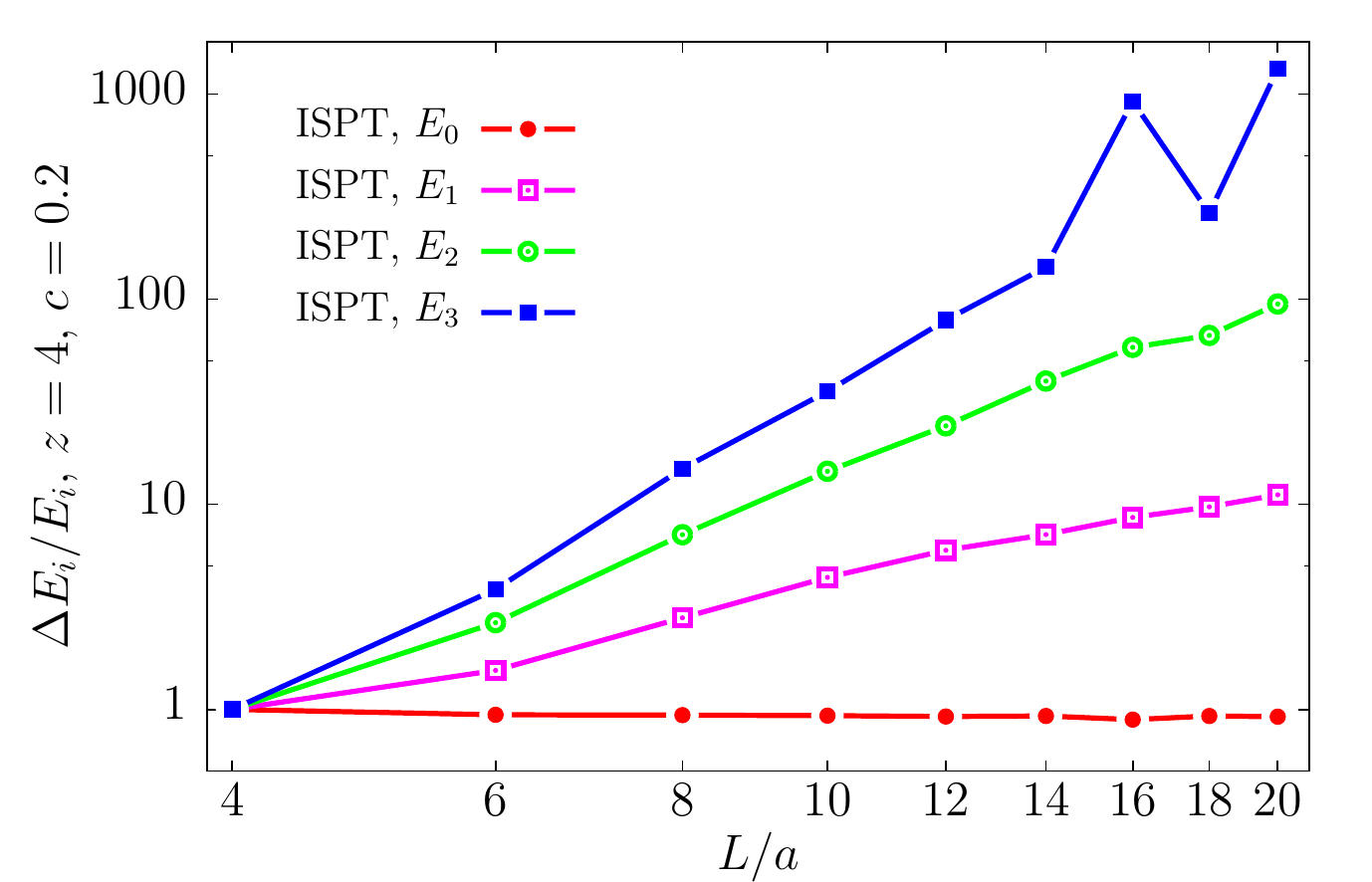}
\caption{Continuum scaling of the relative errors $\Delta E_i/E_i$ of $E_i$ 
as computed with ISPT, LSPT and HSPT (see the text for more details). The case
with $z=4$ and $c=0.2$ is shown.}
\label{fig:RelativeError}
\end{figure}

In Figure \ref{fig:RelativeError} we present the continuum scaling of the relative
errors $\Delta E_i/E_i$ of the perturbative coefficients of $\mathcal{E}$ as obtained with 
ISPT, HSPT and LSPT.\footnote{We note that even though the field and mass are properly renormalized,
the expansion coefficients are in terms of the bare coupling $g_0$. The associated logarithmic 
divergence, however, is not expected to be relevant for our conclusions.} The relative error is 
normalized at its value at $L/a=4$. For the error analysis in HSPT, and LSPT we employed the method
described in~\cite{Wolff:2003sm}. In the case of LSPT we considered two values of the step-size 
$\epsilon$, in order to access the dependence of the results on it. For HSPT, we 
kept the average trajectory length fixed to $\langle\tau\rangle=1$, while decreasing the 
step-size as $(L/a)^{-1/2}$. This was done in order to keep the step-size errors roughly constant
as the continuum limit is approached. Finally, for all methods the number of field configurations
has been kept fixed as $L/a$ is increased, and always separated by a single step (trajectory)
for LSPT (HSPT).

\begin{figure}[!hptb]
\centering
\includegraphics[width=0.50\textwidth]{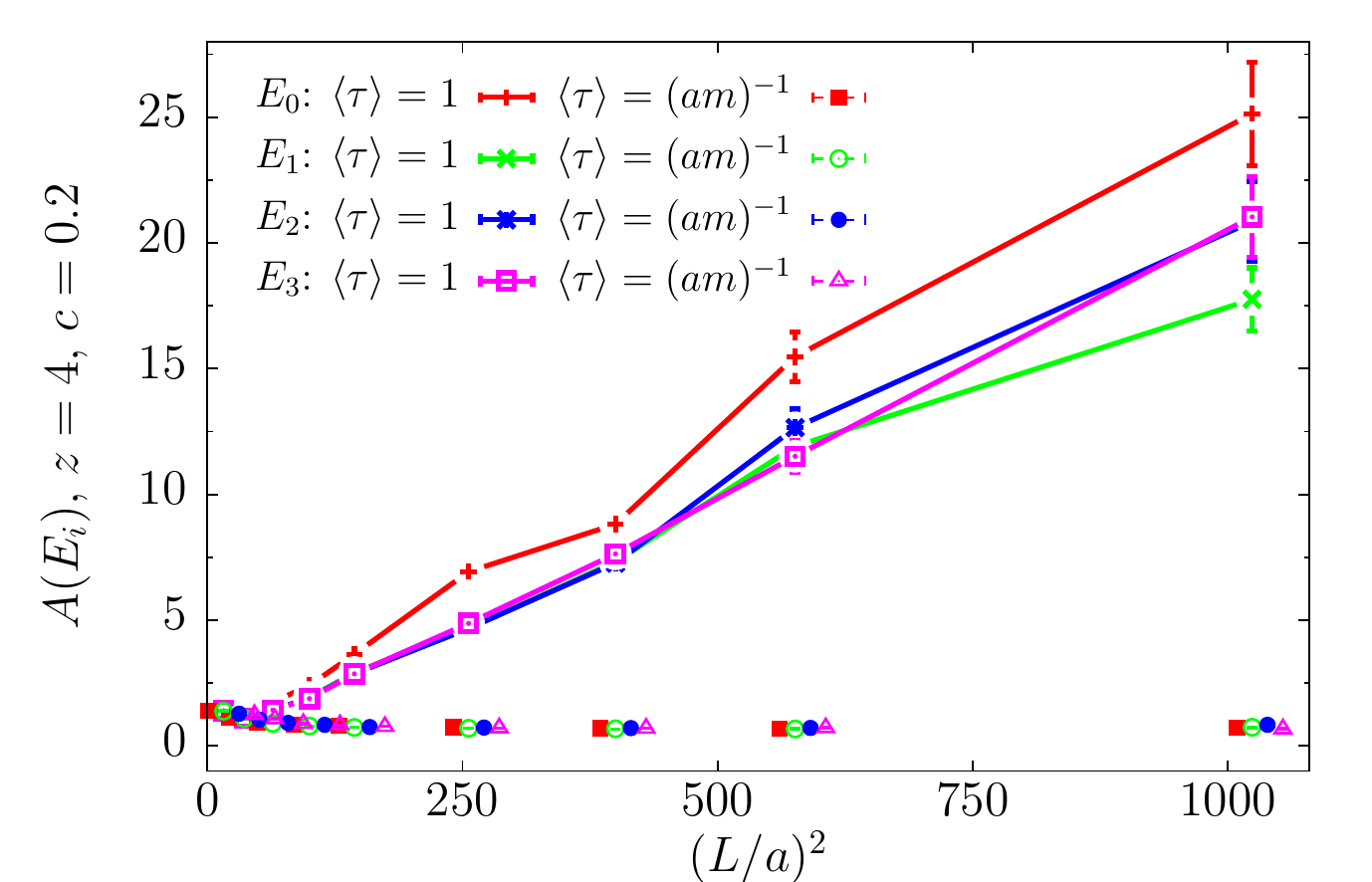}
\caption{Continuum scaling of the integrated autocorrelations $A(E_i)$
of ${E}_i$, $i=0,\ldots3$, for the cases $\langle\tau \rangle=1$ and 
$\langle\tau \rangle=(am)^{-1}$. The points for $\langle\tau \rangle=(am)^{-1}$ 
are shifted along the $x$ axis for clarity.}
\label{fig:Autocorrelations}
\end{figure}

Observe how the relative errors of $\mathcal{E}$ in LSPT and HSPT grow roughly
linearly with $L/a$. This is compatible with the errors growing due to the increase of 
autocorrelations as $(L/a)^2$. ISPT, instead, shows a rather different behavior: the relative
error of the perturbative coefficients $E_i$ grows as increasing powers of $L/a$ as the 
perturbative order is increased. This is seen more clearly in the bottom-right panel of Figure 
\ref{fig:RelativeError}, where a detail of ISPT is shown. Since in ISPT the field configurations
are uncorrelated, this rapid increase in the relative errors of the coefficients must be related
to their variance. This behaviour has been elucidated by L\"uscher~\cite{Luscher:2015ab},
who emphasized the generic presence of power divergences in the variance of perturbative 
coefficients computed with ISPT; he also shows that such power divergences are excluded 
if the fields are generated using the Langevin equations. Finally, we investigated how the integrated
autocorrelation $A(E_i)$ of the perturbative coefficients ${E_i}$ scales with $L/a$ in HSPT. 
In Figure \ref{fig:Autocorrelations} we compare the case $\langle\tau \rangle=1$ with the case
when $\langle\tau\rangle=(am)^{-1}$. For all perturbative orders we considered, the results are 
consistent with the free field theory expectations~\cite{Kennedy:2000ju}, namely $A(E_i)$ for
$\langle \tau \rangle=1$ grows like $(L/a)^2$ whereas for $\langle \tau \rangle=(am)^{-1}$ it 
is constant.

\section{Conclusions}

From this study we conclude that the variance of ISPT grows very rapidly as we increase the
order in $g_0$. Even though it has many appealing features, this technique does not appear
to be competitive in its present form. On the other hand, defining NSPT in terms of different 
stochastic equations, as the HMD or the Kramers equations, is a simple and profitable idea. 
Indeed, this allows us to exploit recent algorithmic advances in the context of stochastic
perturbation theory. This will be extremely useful for more complicated theories such as QCD.

\paragraph{Acknowledgments:} M.D.B. would like to thank Martin L\"uscher for his stimulating
and useful correspondence. Useful discussions with Tomasz Korzec, Christopher Monahan, and 
Ulli Wolff are also greatly acknowledged.

\end{document}